\documentstyle[sprocl]{article}

\def\ra{\rightarrow}

\def\be{\begin{equation}}
\def\ee{\end{equation}}
\def\bea{\begin{eqnarray}}
\def\eea{\end{eqnarray}}

\begin{document}
\def\lta{\;\raisebox{-.5ex}{\rlap{$\sim$}} \raisebox{.5ex}{$<$}\;}
\def\gta{\;\raisebox{-.5ex}{\rlap{$\sim$}} \raisebox{.5ex}{$>$}\;}
\def\grle{\;\raisebox{-.5ex}{\rlap{$<$}}    \raisebox{.5ex}{$>$}\;}
\def\legr{\;\raisebox{-.5ex}{\rlap{$>$}}    \raisebox{.5ex}{$<$}\;}

\newcommand{\beq}{\begin{equation}} 
\newcommand{\eeq}{\end{equation}}
\newcommand{\nl}{\nonumber \\} 

\newcommand{\permille}{$^0 \!\!\!\: / \! _{00}$}
\newcommand{\dd}{{\rm d}}
\newcommand{\oal}{{\cal O}(\alpha)}%
\newcommand{\su}{$ SU(2) \times U(1)\,$}
\newcommand{\TeV}{\mbox{TeV}}
\newcommand{\xdim}{$dim=6$~} 

\newcommand{\eps}{\epsilon}
\newcommand{\mw}{M_{W}}
\newcommand{\mww}{M_{W}^{2}}
\newcommand{\mbb}{m_{b \bar b}}
\newcommand{\mcc}{m_{c \bar c}}
\newcommand{\mbc}{m_{b\bar b(c \bar c)}}
\newcommand{\mh}{m_{H}}
\newcommand{\mhh}{m_{H}^2}
\newcommand{\mz}{M_{Z}}
\newcommand{\mzz}{M_{Z}^{2}}

\newcommand{\lra}{\leftrightarrow}
\newcommand{\tr}{{\rm Tr}}
 
\newcommand{\ie}{{\em i.e.}}
\newcommand{\cm}{{{\cal M}}}
\newcommand{\cl}{{{\cal L}}}
\def\Ww{{\mbox{\boldmath $W$}}}  
\def\B{{\mbox{\boldmath $B$}}}         
\def\nn{\noindent}

\newcommand{\sinsq}{\sin^2\theta}
\newcommand{\cossq}{\cos^2\theta}

\newcommand{\epem}{$e^{+} e^{-}\;$}
\newcommand{\epemt}{e^{+} e^{-}\;}
\newcommand{\eeah}{$e^{+} e^{-} \ra H \gamma \;$}
\newcommand{\eahnw}{$e\gamma \ra H \nu_e W$}

\newcommand{\thebb}{\theta_{b-beam}}
\newcommand{\thebc}{\theta_{b(c)-beam}}
\newcommand{\pte}{p^e_T}
\newcommand{\ptH}{p^H_T}
\newcommand{\gag}{$\gamma \gamma$ }
\newcommand{\gam}{\gamma \gamma }

\newcommand{\aatoh}{$\gamma \gamma \ra H \;$}
\newcommand{\egam}{$e \gamma \;$}
\newcommand{\eat}{e \gamma \;}
\newcommand{\eaeh}{$e \gamma \ra e H\;$}
\newcommand{\eaehb}{$e \gamma \ra e H \ra e (b \bar b)\;$}
\newcommand{\egebb}{$e \gamma (g) \ra e b \bar b\;$}
\newcommand{\egecc}{$e \gamma (g) \ra e c \bar c\;$}
\newcommand{\egebc}{$e \gamma (g) \ra e b \bar b(e c \bar c)\;$}
\newcommand{\eaebb}{$e \gamma \ra e b \bar b\;$}
\newcommand{\eaecc}{$e \gamma \ra e c \bar c\;$}
\newcommand{\aah}{$\gamma \gamma H\;$}
\newcommand{\zah}{$Z \gamma H\;$}
\newcommand{\pe}{P_e}
\newcommand{\pg}{P_{\gamma}}
\newcommand{\delbb}{\Delta m_{b \bar b}}
\newcommand{\delbc}{\Delta m_{b \bar b(c\bar c)}}


\renewcommand{\d}{{\rm d}}
\newcommand{\db}{{\rm d}_{\scriptscriptstyle{\rm B}}}
\newcommand{\bard}{\overline{{\rm d}}}
\newcommand{\bardb}{\overline{{\rm d}}_{\scriptscriptstyle{\rm B}}}
\renewcommand{\aa}{{\rm d}_{\scriptscriptstyle \gamma\gamma}}
\newcommand{\GeV}{\rm GeV}
\newcommand{\az}{{\rm d}_{\scriptscriptstyle \gamma Z}}
\newcommand{\zz}{{\rm d}_{\scriptscriptstyle ZZ}}
\newcommand{\aacp}{\overline{{\rm d}}_{\scriptscriptstyle \gamma\gamma}}
\newcommand{\azcp}{\overline{{\rm d}}_{\scriptscriptstyle \gamma Z}}
\newcommand{\zzcp}{\overline{{\rm d}}_{\scriptscriptstyle ZZ}}
\newcommand{\tgw}{\tan{\theta_W}}
\newcommand{\cotgw}{(\tan{\theta_W})^{-1}}


\begin{center}
{\Large \bf Bounds on anomalous {\boldmath \aah} and {\boldmath \zah}
couplings \\ at future e~{\boldmath $\gamma$} linear colliders}
\end{center}
\bigskip

\begin{center}
{\large 
E.~Gabrielli~$^{1}$ 
\footnote{
Talk given at the {International Workshop on Linear Colliders},
Sitges, Barcellona, Spain, April 28 -- May 5, (1999).},
$\;\,$V.A.~Ilyin~$^2 \;\,$
and $\;\,$B.~Mele~$^3$. 
} \\
\end{center}
\medskip\noindent
$^1$ Departamento de F\'{\i}sica Te\'orica,
        Universidad Aut\'onoma, Madrid, Spain\\
\noindent
$^2$ Institute of Nuclear Physics, Moscow State University, Russia \\
\noindent
$^3$ Physics Department, Roma University 1, and INFN Roma1, Italy

\bigskip
\begin{center}
{\bf Abstract} \\
\end{center}
{\small 
We study the bounds on the anomalous contributions to the
\aah and \zah vertices that can be obtained via the process 
\eaeh. We consider the representative
cases of an intermediate Higgs mass production of 
$\mh =120~\GeV$ and for a center of mass energy of $\sqrt{S}=500~\GeV$
and $\sqrt{S}=1500~\GeV$.
We use a model independent analysis based on \su invariant
operators of \xdim added to the Standard Model lagrangian.
We find that this process provides an excelent way to put strong constraints
both in the sector of CP-even and CP-odd anomalous couplings 
contribution to the \aah and \zah vertices.
}

\section{Introduction}

The Higgs boson sector is a crucial part of the Standard Model (SM) still
escaping direct experimental verification. Once the scalar boson will be
discovered either at LEP2, upgraded TEVATRON or at LHC, testing its properties
will be a central issue at future linear  colliders. In particular, an \epem
collider with centre-of-mass (c.m.) energy $\sqrt{s}\simeq (300\div 2000)$GeV
and integrated luminosity ${\cal O}(100)$ fb$^{-1}$ will allow an accurate
determination of the mass, some couplings and  parity properties of this new
boson \cite{saar,zerw}. Among other couplings, the interaction of scalars  
with
the neutral electroweak gauge bosons,  $\gamma$ and $Z$, are particularly
interesting. Indeed, one can hope to test here some delicate feature of the
Standard Model --- the relation between the spontaneous symmetry breaking
mechanism and the electroweak mixing of the two gauge groups $SU(2)$ and
$U(1)$. In this respect, three vertices could be measured --  $ZZH$,
$\gamma\gamma H$ and $Z\gamma H$. While the $ZZH$ vertex stands in SM at the
tree level, the other two contribute only at one-loop. This means that  the
\aah and \zah couplings could be sensitive to the contributions of new
particles circulating in the loop.

Here, we discuss the case of  an intermediate-mass Higgs boson, that is with
$M_Z\lta \mh\lta 140$ GeV.  A measurement of the \aah coupling should be
possible by the determination of the BR for the decay $H\ra \gamma \gamma$,
e.g. in the LHC Higgs discovery channel, $gg\to H\to\gamma\gamma$. 
Furthermore, at future photon-photon colliders\footnote{Two  further options
are presently considered for a high-energy $e^+e^-$ linear collider,  where 
one
or both the initial $e^+/e^-$ beams are replaced by   photon beams induced by
Compton backscattering of laser light on the high-energy electron beams
\cite{spec}. Then, the initial real photons could be to a good degree
monochromatic, and have energy and luminosity comparable to the ones of the
parent electron beam \cite{mono}.}, the precise measurement of the  \aah 
vertex looks  realistic at the resonant production of the Higgs particle,
$\gamma\gamma\to H$. To this end, the capability of tuning the \gag c.m. 
energy on the Higgs mass, 
through  a good degree of the photons  monochromaticity,
will be crucial for not diluting too much  the $\gam \to H$  resonant cross
section over the c.m. energy spectrum. Measuring the  \zah vertex is in 
general
more complicated.  Indeed, if one discusses the corresponding Higgs decay, the
final states include the $Z$ decay products, jets or lepton pairs, where much
heavier backgrounds are expected. Then, one can discuss  the $H\to \gamma Z$ 
decay only for $m_H\gta 115$ GeV (and $m_H\lta 140$),  when the corresponding
branching is as large as ${\cal O}(10^{-3})$.  

Another possibility of measuring
the \zah vertex  is given by  collision processes. At electron-positron
colliders, the corresponding channels are $ e^+e^- \to \gamma H$ and $e^+e^-
\to Z H$.  However, in the $ZH$ channel the \zah vertex contributes to the
corresponding  one-loop corrections, thus implying a large tree level
background. The reaction $e^+e^-\to \gamma H$ has been extensively studied in
the literature \cite{barr,abba,djou}. Unfortunately, the \eeah channel suffers
from small rates, which are further depleted at large energies by the $1/s$
behavior of the dominant s-channel diagrams. For example, $\sigma_S\approx
0.05\div 0.001$ fb at $\sqrt{s}\sim 500\div 1500$ GeV.  We estimated the  main
background coming from the $e^+e^-\to\gamma b\bar b$ process, and found it
rather heavy: $\sigma_B\approx 4\div 0.8$ fb for $m_{b\bar b}=100\div 140$ 
GeV,
assuming a high resolution in the measurement of the  invariant  mass of 
$b\bar
b$ quark pair, i.e. $\pm 3$ GeV, and applying a minimum  cut of $18^\circ$  on
the angles [$\gamma-beams$] and [$b(\bar b)-beams$].   Then, at $\sqrt{s}=1.5$
TeV, we get $\sigma_B\approx 0.4\div 0.07$ fb. One can conclude that
measuring the \zah vertex is not an easy task.  

Recently, the Higgs production
in electron-photon collisions through the one-loop process   $e\gamma \to eH$
was analysed in details  \cite{noii}. This channel will turn out to be an
excellent tool to test  both the \aah and \zah one-loop couplings with high
statistics, without requiring a fine tuning of the c.m. energy. 

In this paper we analyse the prospects of the \eaeh reaction in setting
experimental bounds on the value of the anomalous \aah and \zah couplings
\cite{paper}. 
For this analysis we use a model independent approach, 
where \xdim \su invariant operators are added to the SM Lagrangian. 
In realistic  models extending the 
SM, these operators contribute in some definite combinations. However, if one
discusses the bounds on the possible deviations from  the standard-model
one-loop Higgs vertices, this approach can give some general  insight into the
problem. These anomalous operators contribute to all the three vertices   
\aah, \zah and $ZZH$, with only the first two involved in the \eaeh reaction.  
Even though the anomalous contributions to the \aah vertex 
can be bounded through the resonant $\gamma\gamma\to H$ reaction, 
competitive bounds can be obtained by measuring the total rate 
of the discussed reaction, \eaeh. 

The paper is organized as follows:
in the next section we present the \eaeh process
and outline its main features.
In section 3 we present the \xdim operators which induces
the anomalous vertices contributions to the \eaeh amplitude.
Section 4 contains the numerical results 
for the bounds on the anomalous couplings and the  conclusions.
\section{The reaction \eaeh in the SM: main features}

In \cite{noii}, we presented the complete analytical results for the helicity 
amplitudes of the \eaeh process in the SM (see also reference \cite{cott}). 
This amplitude is given by the diagrams contribution denoted as 
`\aah' and `\zah', which are related to the \aah and \zah vertices 
respectively, and a 'BOX' contribution. 
The separation of the rate into these three parts corresponds 
to the case  where the Slavnov-Taylor identities for the `\aah' 
and `\zah' Green  functions just imply the transversality with respect to 
the incoming photon momentum.

The total rate of this reaction is rather high, in particular
for $\mh$ up to about 400 GeV, one finds $\sigma>1$ fb. 
If no kinematical cuts are imposed then
the main contribution to the cross section is given by the \aah vertex;
this is due to the t-channel photon propagator given by the
\aah vertex. 
On the contrary the \zah vertex contribution 
is depleted by the $Z$ propagator. 

Nevertheless, as discussed in  
\cite{noii}, the \zah vertex effects can be 
extracted from \eaeh by implementing a suitable strategy to reduce the 
\aah vertex contribution:
this require a final electron tagged at large angle together with 
a transverse momentum cut $\pte>100$ GeV.
For example, for $\pte>100$ GeV, we found that \zah is about 60\% of \aah, and 
\zah gives a considerable fraction of the total production rate, 
which is still sufficient to guarantee investigation (about 0.7 fb).

The main irreducible background to the process $e\gamma\to eH\to eb\bar b$ 
comes from the channel \eaebb.  A further source of background is the charm
production through \eaecc, when the $c$ quarks are misidentified into $b$'s.
We also assume a 10\% probability  of misidentifying a  $c$
quark into a $b$.
The cut $\thebc>18^\circ$ (between each $b(c)$ quark and both the
beams reduces the signal and background at a comparable level.  
Numerically the \eaecc ``effective rate" is of the same order as
the \eaebb rate. 
A further background, considered in  
\cite{noii}, is the resolved \egebc
production, where the photon interacts via its gluonic content.
This background was found negligible.

Important improvements in the $S/B$ ratio can be obtained by exploiting the
final-electron angular asymmetry in the signal. Indeed,  the final electron in
\eaeh moves mostly in the forward direction.    

The main conclusion obtained in \cite{noii} is the following: with a
luminosity of 100 fb$^{-1}$, at $\sqrt{s}=500$GeV, one expects an accuracy as
good as about 10\% on the measurement of the \zah effects assuming the 
validity of Standard Model. Therefore we can use the suitable strategy used in
\cite{noii} to study the sensitivity of the \eaeh process 
to the anomalous coupling contributions in both the \aah and \zah vertices.

\section{Anomalous vertices}

Now we consider the possibility that the new physics affects 
the bosonic sector of the SM through low energy effective operators of 
\xdim. These operators contribute to the \eaeh amplitude 
via anomalous couplings to the \aah and \zah vertices.

In particular there are two pairs of \xdim 
operators\footnote{We assume that $SU(2)\times U(1)$ 
local gauge invariance of the Standard Model should be valid as well as
so-called custodial symmetry of the gauge and Higgs sectors present in the
Standard Model \cite{anomOP}}, 
CP-even and CP-odd respectively, giving anomalous contributions 
to the process \eaeh:

\beq
{\cal L}^{eff} = \d\cdot {\cal O}_{UW} + \db\cdot {\cal O}_{UB} +
     \bard\cdot \bar {\cal O}_{UW} 
                        + \bardb\cdot \bar {\cal O}_{UB},
\eeq

\beq
{\cal O}_{UW} \,=\, \frac{1}{v^2} \left( |\Phi|^2-\frac{v^2}{2}\right)
    \cdot W^{i\mu\nu}  W^i_{\mu\nu}, \qquad
   {\cal O}_{UB} \,=\, \frac{1}{v^2} \left( |\Phi|^2-\frac{v^2}{2}\right)
    \cdot B^{\mu\nu}  B_{\mu\nu},
\eeq

\beq
\bar {\cal O}_{UW} \,=\, \frac{1}{v^2}  |\Phi|^2
    \cdot W^{i\mu\nu}  \tilde W^i_{\mu\nu}, \qquad
   \bar {\cal O}_{UB} \,=\, \frac{1}{v^2} |\Phi|^2
    \cdot B^{\mu\nu}  \tilde B_{\mu\nu}, 
\eeq
where
$ \tilde W^i_{\mu\nu} = \epsilon_{\mu\nu\mu'\nu'}\cdot W^{i\mu'\nu'}$ and
$ \tilde B_{\mu\nu} = \epsilon_{\mu\nu\mu'\nu'}\cdot B^{\mu'\nu'}$.
In these formulas $\Phi$ is the Higgs doublet and 
$v$ is the electroweak vacuum expectation value.

The \aah and \zah anomalous terms contribution, in terms of
$\d,\db,\bard\;,\bardb $ couplings, to 
the helicity amplitudes of \eaeh can be found in \cite{paper}.
\section{Bounds on anomalous \aah and \zah couplings:
numerical results and conclusions}

In this section we analyse the numerical results \cite{paper}
for the bounds on the anomalous couplings $\d,~\db$ and $\bard,~\bardb$
which are obtained from the \eaeh process.
\footnote{
Most of the results presented in this work were obtained with the help 
of CompHEP package \cite{comp}.} 
These bounds have been computed by using the
requirement that no deviation from the SM cross section is observed at the 
95\% CL, in particular we have:
\beq
N^{\rm anom}(\kappa) < 1.96 \cdot \sqrt{N^{\rm tot}(\kappa)}, \quad 
   \kappa = \d,\db,\bard\;,\bardb\;, 
\eeq
\beq
N^{\rm tot}(\kappa) = {\cal L}_{int} \cdot [\sigma_S(\kappa)+\sigma_B]\;,
  \quad
   N^{\rm anom}(\kappa) = {\cal L}_{int} \cdot 
     [\sigma_S(\kappa)-\sigma_S(0)]\;.
\eeq
where ${\cal L}_{int}$ is the integrated luminosity, 
$N^{\rm tot}$ and $N^{\rm anom}$ denote respectively 
the total number of observed events and
the anomalous number of events deviating from the expected SM predictions
for the signal.
Here, by $\sigma_S(\kappa)$ we mean the cross  section of the signal 
reaction $e\gamma\to eH\to eb\bar b$ with the anomalous contributions, so  
$\sigma_S(0)$ is the SM cross section.  Then, by $\sigma_B$ we denote the total
cross section of the  background processes  \eaebb, \eaecc (with 10\%
probability of misidentifying a $c$ quark into a $b$ quark).

The complete numerical results for the bounds on the 
CP-even $\d,~\db$ and CP-odd $\bard\;,\bardb$  
anomalous couplings can be found in \cite{paper}.
They are all obtained for a representative case of 
$\mh=120~\GeV$, and for $\sqrt{s}=500~\GeV$ and $\sqrt{s}=1500~\GeV$, 
and for different electron beam polarizations 
$P_e=0,1,-1$, attainable with an integrated luminosity of 100 fb$^{-1}$
and 1000 fb$^{-1}$ respectively.
In all these results we always assume that for each bound 
the only contribution is given by the corresponding anomalous coupling,
switching-off the other three anomalous contributions. 

From the results of reference \cite{paper}
we draw the following conclusions:
\begin{itemize}
\item The strongest bounds on the CP-even couplings at $\sqrt{s}=500$~GeV
 are at the level of
$|\d| \lta 6 \times 10^{-4}$, obtained at $\pe=-1$, 
and $|\db| \lta 2.5 \times 10^{-4}$
(with no cut on $\pte$), not depending on the $e$ polarization.
At $\sqrt{s}=1500$~GeV, one has 
$|\d| \lta 1.7 \times 10^{-4}$, obtained at $\pe=-1$ and $\pte >100$~GeV, 
and $|\db| \lta 1 \times 10^{-4}$
(with no cut on $\pte$), not depending on the $e$ polarization.
\item The strongest bounds on the CP-odd couplings at $\sqrt{s}=500$~GeV
are $|\bard| \lta 3 \times 10^{-3}$ and $|\bardb| \lta 1 \times  10^{-3}$, 
which are obtained for $\pe=-1$ and
$\pe=1$, respectively. These bounds are quite 
insensitive to the cuts on $\pte$.
At $\sqrt{s}=1500$~GeV, one has $|\bard| \lta 1.0 \times 10^{-3}$ 
for $\pe=-1$, with $\pte>100$~GeV, and
$|\bardb| \lta 3 \times  10^{-4}$, for $\pe=1$, with $\pte>100$~GeV.
\end{itemize}

Other processes have been studied in the literature that could be able to 
bound the parameters $\d,~\db,~\bard,~\bardb$ at future linear colliders. 
In particular, the processes $e^+ e^- \to HZ$ and $\gamma \gamma \to H$
have been studied for a \epem collider at $\sqrt{s}=1$~TeV and 
with 80 fb$^{-1}$ by Gounaris et al.
From $e^+ e^- \to HZ$, they get 
$|\d| \lta 5 \times 10^{-3}$,
$|\db| \lta 2.5 \times 10^{-3}$, $|\bard| \lta 5 \times 10^{-3}$
and $|\bardb| \lta 2.5 \times 10^{-3}$  \cite{eeHZ}.
The process $\gamma \gamma \to H$ can do a bit better and reach the values
$|\d| \lta 1 \times 10^{-3}$,
$|\db| \lta 3 \times 10^{-4}$, $|\bard| \lta 4 \times 10^{-3}$
and $|\bardb| \lta 1.3 \times 10^{-3}$, assuming a particular photon
energy spectrum \cite{gouna96}.
These analysis assume a precision of the measured 
production rate equal to $1/\sqrt{N}$ (with $N$ the total number of
events), and neglect possible backgrounds.
\noindent
In order to set the comparative potential of our process with respect
to these two processes in bounding the parameters  
$\d,~\db,~\bard,~\bardb$,  we assumed $\sqrt{s}=0.9$~TeV and (conservatively)
a luminosity of  25 fb$^{-1}$ in  \eaeh.
We then neglected any background, and assumed a precision 
equal to $1/\sqrt{N}$. 
In the case $\pe=0$ and $\pte >0$, we get 
$|\d| \lta 5 \times 10^{-4}$,
$|\db| \lta 2 \times 10^{-4}$, $|\bard| \lta 2 \times 10^{-3}$
and $|\bardb| \lta 8 \times 10^{-4}$.

\noindent
This analysis confirms the excellent potential of the process \eaeh.

Following the conventions of reference \cite{9811413},
one can convert these constrains into upper limits of the {\it new physics}
scale $\Lambda$ that can be explored through 
\eaeh with $\sqrt{s}\simeq 1.5$~TeV and 10$^3$ fb$^{-1}$:
\begin{equation}
\begin{array}{rcl}
|\d| \lta 1.7 \times 10^{-4}     
& \to &    |\frac{f_{WW}}{\Lambda^2}| \lta  0.026  ~~ \TeV^{-2} \\\\
|\db| \lta 1.0 \times 10^{-4}     
& \to &    |\frac{f_{BB}}{\Lambda^2}| \lta  0.015  ~~ \TeV^{-2} \\\\
|\bard| \lta 1.0 \times 10^{-3}     
& \to &   |\frac{\bar f_{WW}}{\Lambda^2}| \lta  0.15 ~~  \TeV^{-2} \\\\
|\bardb| \lta 3.0 \times 10^{-4}     
& \to &   |\frac{\bar f_{WW}}{\Lambda^2}| \lta  0.046  ~~  \TeV^{-2} 
\end{array}
\end{equation}
\noindent
For $f_i\sim 1$ one can explore energy scales up to about 
6, 8, 2.6 and about 4.5 TeV, respectively. 
At $\sqrt{s} \simeq$ 500 GeV and $10^2$ fb$^{-1}$, the 
corresponding contraints on the couplings are a factor 2 or 3 weaker
than above
(reflecting into energy scales $\Lambda$ lower by a factor 1.4 or 1.7,
respectively),
mainly because of the smaller integrated luminosity assumed.

\section*{Acknowledgements}
E.G. acknowledges the financial support of the TMR network
project ref. FMRX-CT96-0090 and partial financial support from the 
CICYT project ref. AEN97-1678.
V.I. was partly supported by the joint RFBR-DFG grant 99-02-04011.

\end{document}